\documentclass[conf]{new-aiaa}
\usepackage[utf8]{inputenc}
\usepackage{multicol}
\usepackage{graphicx}
\usepackage{amsmath}
\usepackage[version=4]{mhchem}
\usepackage{siunitx}
\usepackage{longtable,tabularx}
\usepackage{wrapfig} 
\usepackage{comment}  
\usepackage{subcaption} 
\usepackage{float}
\usepackage{booktabs}
\title{Flow-Driven Rotor Simulations of Seyi-Chunlei Ducted Turbine}

\author{Seyi Oluwadare\footnote{Graduate Student, AIAA Student Member} and Chunlei Liang \footnote{Professor, AIAA Associate Fellow}}
\affil{Department Mechanical and Aerospace Engineering, Clarkson University, Potsdam, NY 13699}

\begin{document}

\maketitle

\begin{abstract}
This paper proposes an improved Clarkson Ducted Wind Turbine (DWT) design using a new diffuser based on the Selig S1223 airfoil at an angle of attack (AoA) of 20 degrees and a smaller tip clearance. This proposed design is hereby named Selig20 Clarkson Ducted Turbine or Seyi-Chunlei Ducted Turbine (SCDT) compared to the original Clarkson Ducted Wind Turbine (CDWT). In both SCDT and CDWT configurations, the rotor is placed a distance behind the throat of the duct. For in-depth analysis, we employ a flow-driven-rotor (FDR) model of a commercial CFD package, Simerics-MP+, based on unstructured-grid finite-volume solutions of Unsteady Reynolds-Averaged Navier-Stokes (URANS) equations for the flow field that are two-way fully coupled with a dynamic solution of the rigid-body rotation of the turbine rotor. The FDR model successfully predicts the optimal thrust coefficient, whereas the prescribed rotation model fails to do so. Although the optimal Cp predicted by the FDR model is fairly close to the prediction from the prescribed motion model, FDR is generally more accurate in predicting underperformance under ambient wind conditions away from the optimal tip speed ratio. FDR offers a new path to simulate ducted wind turbines in ambient wind conditions. The Seyi-Chunlei Ducted Turbine is confirmed to have a Cpt peak approximately 7\% higher than that of the Clarkson DWT. SCDT also has a wider range of optimal tip speed ratios, enabling it to harvest more wind energy under ambient conditions. 

\end{abstract}

\section*{Nomenclature}

\begin{multicols}{2}
\noindent
\begin{tabular}[t]{@{}l @{\quad=\quad} l@{}}
c   & chord \\
$h$  & height \\
d$t$ & time step \\
$\lambda, TSR$ & tip speed ratio \\
$\omega_{rotor}, \dot{\theta}$ & rotor angular speed \\
$\alpha, \dot{\omega}$ & angular acceleration \\
$\theta$ & angular displacement \\
$\tau$ & torque \\
$U_{\infty}$ & freestream wind speed \\
$A_{rot}$ & rotor area\\
$A_{exit}$ & duct exit area\\
$D$ & rotor diameter \\
$D_{hub}$ & hub diameter \\

\end{tabular}

\begin{tabular}[t]{@{}l @{\quad=\quad} l@{}}
$R_{in}$ & duct inlet radius \\
$R_{exit}$ & duct exit radius \\
$R_{rot}$ & rotor radius, also named as $R_b$\\
$\Delta r$ & tip clearance \\
$Z_{rot}$ & rotor location from duct inlet \\
$I_{rot}$ & moment of inertia of the rotor \\

$C_P$& power coefficient \\
$C_{pt}$& maximum power coefficient per device area \\
$C_T$& thrust coefficient \\
$DWT$ & ducted wind turbine \\
$MGI$ & mismatched grid interface \\
$P, T$ & power  and thrust of the DWT respectively \\

\end{tabular}
\end{multicols}

\newpage

\section{Introduction}
\label{sec:intro}

According to the International Energy Agency, wind power represented 4. 2\% of the total energy generated in 2017 and is expected to reach a quarter of all energy generation by 2040 \cite{ITE2019}. Competitive ducted wind turbines (DWT) require consistently higher power output under transient wind conditions. DWTs facilitate an increase in the mass flow rate locally. They are insensitive to turbulent fluctuations and yawed flow conditions compared to open rotors. Competitive DWTs represent a new way to source energy at the local level, helping to develop community-based energy distribution and consumption models.

\begin{figure}[ht!]
\centering
\includegraphics[width=14cm]{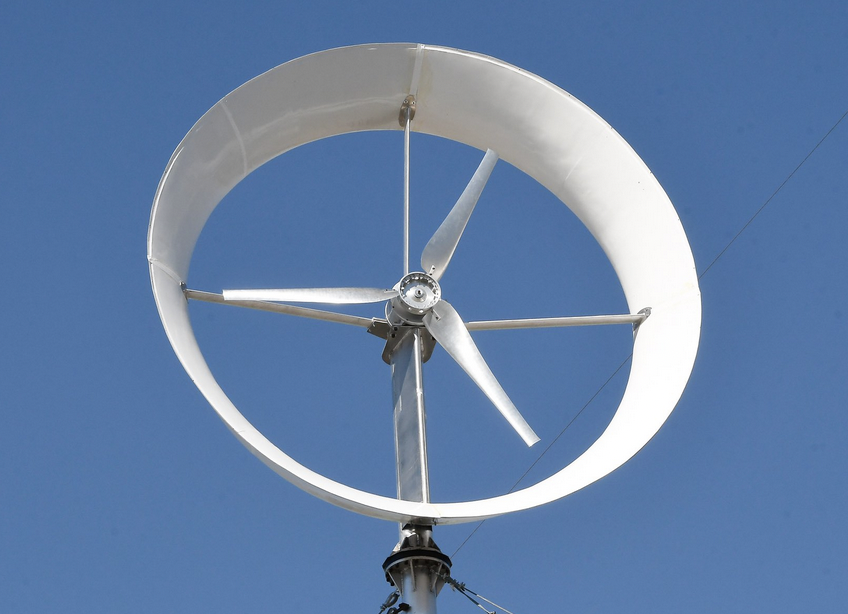}
\caption{Clarkson Generation I Three-Blade Ducted Wind Turbine (CDWT)}
\label{DWT3blade}
\end{figure}

Clarkson University has been developing a full-scale 3.5 kW ducted wind turbine since 2018 and recently installed a unit on a 12-meter monopole tower at the Clarkson University turbine test site.
The ducted wind turbine designed and built at Clarkson University is unique due to its ability to harness wind energy more effectively through aerodynamic optimization guided by Reynolds-Averaged Navier-Stokes (RANS) and actuator disk models \cite{Bagheri2018,Bagheri2021,Visser2024}. Like other diffuser-augmented wind turbines, the Clarkson design is based on increased wind velocity and higher power output from relatively low wind speeds, as shown in Figure \ref{DWT3blade}. 

The Clarkson ducted wind turbines have a rotor diameter D of 3.0m and have performed in ambient conditions considerably below the initial wind tunnel data \cite{Visser2024}.  These ducted wind turbines used the Eppler 423 airfoil at an angle of attack (AoA) of 25 degrees for the diffuser according to an experimental study \cite{Kanya2018}. Rotor blades were created based on GOE417a cambered plate airfoils \cite{Kummer2020} which work very well for small wind turbines, as shown in Figure \ref{DWT3blade}. These airfoils can substantially reduce the blade production cost as a result of their low tooling and manufacturing costs. Their hub diameter is 0.155D according to Safford et al. \cite{Safford2024}.

 However, according to Visser \cite{Visser2024}, the measured Cpt values have exhibited statistically averaged values significantly below a conservative target Cpt value of 0.41 that is $0.5\times 0.9 \times 0.9$ where $Cpt=0.5$ is quantified by measuring a 2.5 meter Clarkson Ducted Wind Turbine (CDWT) in a wind tunnel. This paper improves the CDWT design using a diffuser based on the Selig S1223 airfoil at an angle of attack (AoA) of 20 degrees. This improved design is hereby named Selig20 Clarkson or Seyi-Chunlei Ducted Turbine (SCDT).

The aerodynamic performances of both SCDT and CDWT configurations are investigated thoroughly in this paper using flow-driven-rotor (FDR) URANS simulations on massively parallel computers at Clarkson University. To appreciate the importance of a high-fidelity computational model, it is noted that the present study employs a more sophisticated model, as discussed in Section \ref{computermodel}, than the prescribed rotation model adopted by a recent study \cite{Safford2024} using the same software license. The results in this paper demonstrate the necessity of adopting the flow-driven-rotor model for diffuser optimization, especially when tip speed ratios are non-optimal. To the best knowledge of the authors, this paper is the first to report flow-driven rotor simulation results for a ducted wind turbine. The optimization strategy of this paper is not based on the aerodynamic performance of an individual component of the ducted wind turbine but based on the entire system, that is, the unsteady aerodynamic interaction between the rotor, the duct and the hub. 

\section{Design Parameters of Seyi-Chunlei Ducted Turbine}

\begin{figure}[ht!]
\centering
\includegraphics[width=10cm]{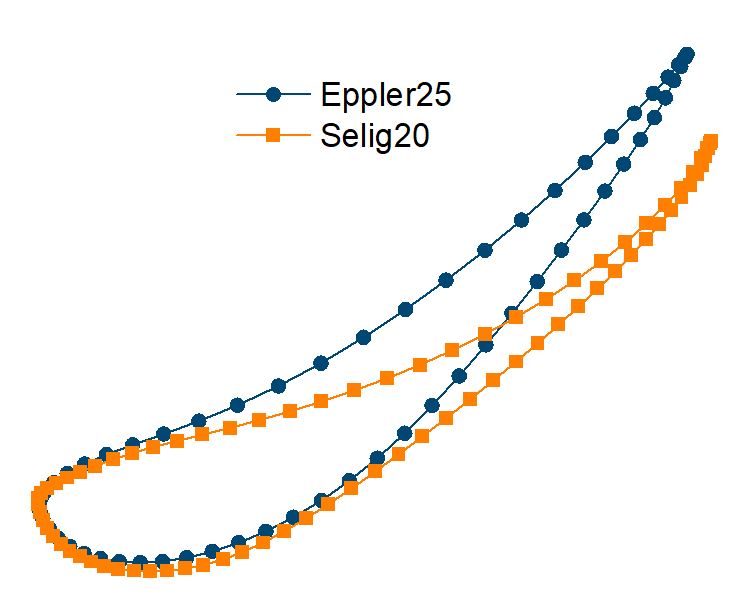}
\caption{Different diffuser designs of SCDT using the Selig 1223 airfoil at $20^{o}$ AoA (Selig20) and CDWT using the Eppler 423 airfoil at $25^{o}$ AoA (Eppler25).}
\label{Selig20_Eppler25}
\end{figure}

Throughout this paper, the diameter of the rotor, D, is kept at 3.0 m.  As mentioned above, SCDT is equipped with a new diffuser based on the Selig S1223 airfoil at an angle of attack of 20 degrees. The Selig airfoil at AoA of 20 degrees is plotted in Figure \ref{Selig20_Eppler25} compared to the Eppler airfoil at AoA of 25 degrees. 

Both airfoils have a chord length of $0.225D$. The Eppler 423 airfoil is known to have a high lift and the airfoil has a thick trailing edge and is easier to construct. At zero angle of attack, the Selig 1223 airfoil is expected to have a higher lift coefficient than the Eppler 423 airfoil. Selig20 gives a smaller exit area of the diffuser than Eppler25.

\newpage

\begin{wrapfigure}{r}{0.4\textwidth} 
    \centering
    \includegraphics[width=\linewidth]{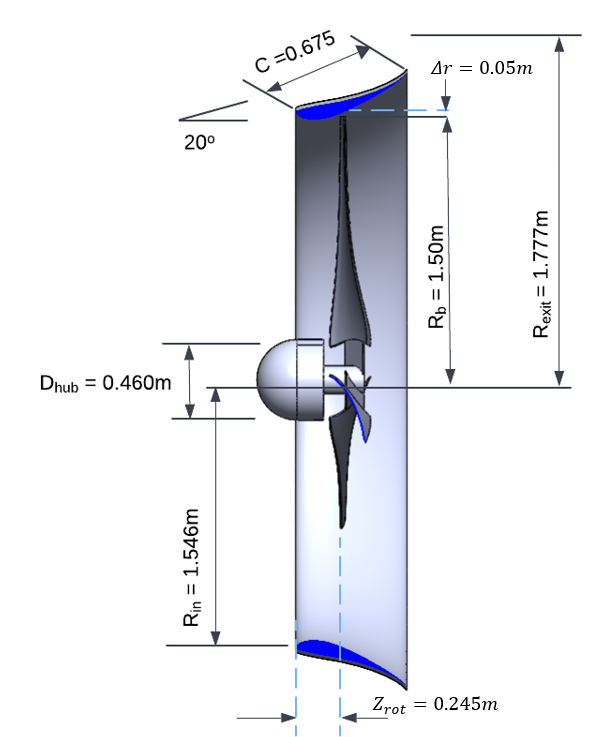}
    \caption{Critical design parameters for the Seyi-Chunlei Ducted Turbine}
    \label{SCDT_parameters}
\end{wrapfigure}

The current SCDT design adopts a three-blade rotor shown in Figure \ref{SCDT_parameters}.  These rotor blades are based on the GOE417a cambered plate airfoils \cite{Kummer2020}. The tip clearance $\Delta r =0.05m$ is about half that of the Clarkson DWT. In both SCDT and CDWT configurations, the rotor is placed a distance behind the throat of the duct (rotor-at-rear). For SCDT, the rotor is placed $0.245m$ behind the inlet of the diffuser. The diameter of the exit of the diffuser is $1.185D$.

\begin{wrapfigure}{r}{0.4\textwidth} 
    \centering
    \includegraphics[width=\linewidth]{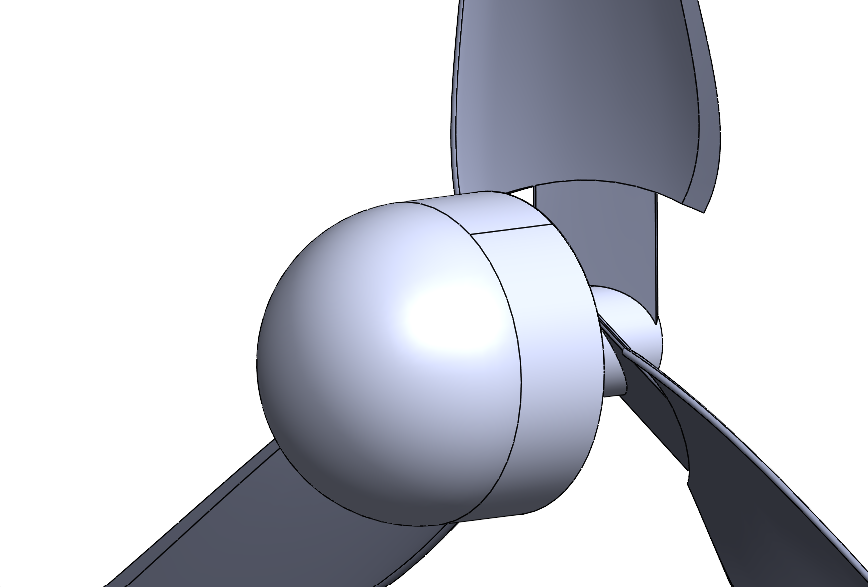}
    \caption{A revised hub configuration for the present computational study}
    \label{hubrevision}
\end{wrapfigure}

As shown in Figure \ref{hubrevision}, SCDT is also installed with a revised hub compared to the hubs adopted by Ding et al. \cite{Ding2023} and Safford et al. \cite{Safford2024}. The frontal hemisphere has a shape identical to the previous designs. The diameter of the frontal hub is 0.155D. However, the rear hub hemisphere is trimmed as shown in Figure \ref{hubrevision} to reduce the generation of hub vortices.

\section{Computational Model}
\label{computermodel}
\subsection{The FDR model}

The mismatched grid interface (MGI) algorithm in SimericsMP+ is a robust feature designed to simulate fluid flows involving moving boundaries and deformable domains, making it essential for applications such as rotating machinery. The present study employs a flow-driven-rotor (FDR) model of a commercial CFD package, Simerics-MP+, based on unstructured-grid finite-volume solutions of URANS equations for the flow field that are two-way fully coupled with a dynamic solution of the rigid-body rotation of the turbine rotor.  The solution starts with the fluid solver by solving three-dimensional incompressible Navier-Stokes equations with standard $k-\epsilon$ turbulence models to compute a fluid torque $\tau_{fluid}$ exerted by the fluid to the rotor. The second-order implicit Euler scheme was used for the time-stepping of the momentum equations. Simerics-MP+ uses pressure-velocity coupling techniques similar to those of PISO and SIMPLE to solve incompressible viscous flows, ensuring that pressure and velocity fields are consistently updated to satisfy mass conservation and momentum balance.  Subsequently, the fluid torque $\tau_{fluid}$ is used as input to a rigid body-rotation dynamics solver according to Equation (\ref{eq:rigid_body}) to calculate the rotor’s angular speed $\omega_{rotor}$, using an explicit time-marching scheme. $\omega_{rotor}$ is then fed into the fluid solver to determine the rotational rate of the spinning domain and the boundary conditions of the moving wall. Finally, the results of the FDR simulations are compared to those of the prescribed rotation model described in \cite{Safford2024,Ding2023} to evaluate differences in flow characteristics, rotor dynamics, and turbine efficiency. 

In the FDR model, the initial angular position and the initial velocity were set to zero, so the rotor starts from rest. The spinning motion along the central axis is perfectly constrained to one degree of freedom. The governing equation for force balance is derived from Newton's Second Law of angular momentum:
        \begin{equation}
            \tau_{fluid}=I_{rot}\alpha +C_{d}\dot{\theta}+\tau_{preload}
            \label{eq:rigid_body}
        \end{equation}

where $\tau_{net}=\tau_{fluid}-C_{d}\dot{\theta}-\tau_{preload}$ is the net torque acting on the system, $I$ is the moment of inertia for the rotor, and $\alpha=\dot{\omega}_{rot}$ is the angular acceleration of the rotor. $\tau_{preload}$ is a preload torque which is typically an intentional application of a prescribed torque to a turbine system to achieve a specific initial tension or force on the components. This preload torque helps the simulation to reach a steady state and converge quickly. $\tau_{damping}= C_{d}\cdot\dot{\theta}$, accounts for energy dissipation, such as rotational friction or aerodynamic drag. In this paper, the damping coefficient $C_{d}$ is varied to obtain a target tip speed ratio $\lambda = \frac{\omega_{rot}R_{rot}}{U_{\infty}}$ when the net torque vanishes. 

\begin{figure}[hbt!]
    \centering
    \includegraphics[width=10cm]{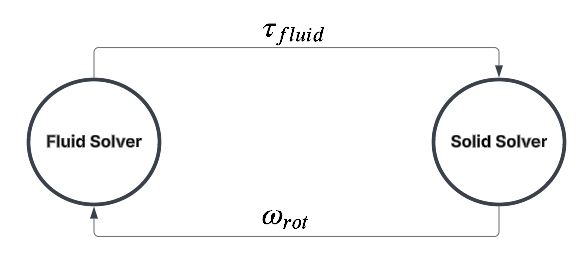}
    \caption{FDR interaction at the interface between the fluid solver and solid solver}
    \label{FDRsolver}
\end{figure}

The interface facilitating a strong two-way coupling between the fluid solver and the rigid solver is shown in Figure \ref{FDRsolver}. The process starts from solving the Navier-Stokes equations using a moving wall boundary condition determined by the rotational rate of the blades to obtain the pressure and shear stress distribution on the rotor to obtain the fluid torque. The fluid torque $\tau_{fluid}$ is then used as an input in the rigid body solver to calculate the angular speed of the rotor $\omega_{rot}$.

\subsection{Computational Setup}

For both CDWT and SDCT configurations, the moment of inertia of the rotor is $I_{rot} =18.09 kg\cdot m^{2}$. All simulations are performed at a Reynolds number of 2.05 million based on the rotor diameter and the free-stream velocity. Most simulations for this paper were conducted on Clarkson's ACRES supercomputer. Each unsteady dynamic-mesh simulation runs on either 80 or 120 nodes and takes approximately 3 days on a coarse mesh with a total of 13 million cells.  The fluid solution was updated by a second-order implicit time-stepping, and an explicit time discretization method was used in the rigid body solver.

\subsubsection{Boundary Condition}

\begin{figure}[hbt!]
    \centering
    \includegraphics[width=11cm]{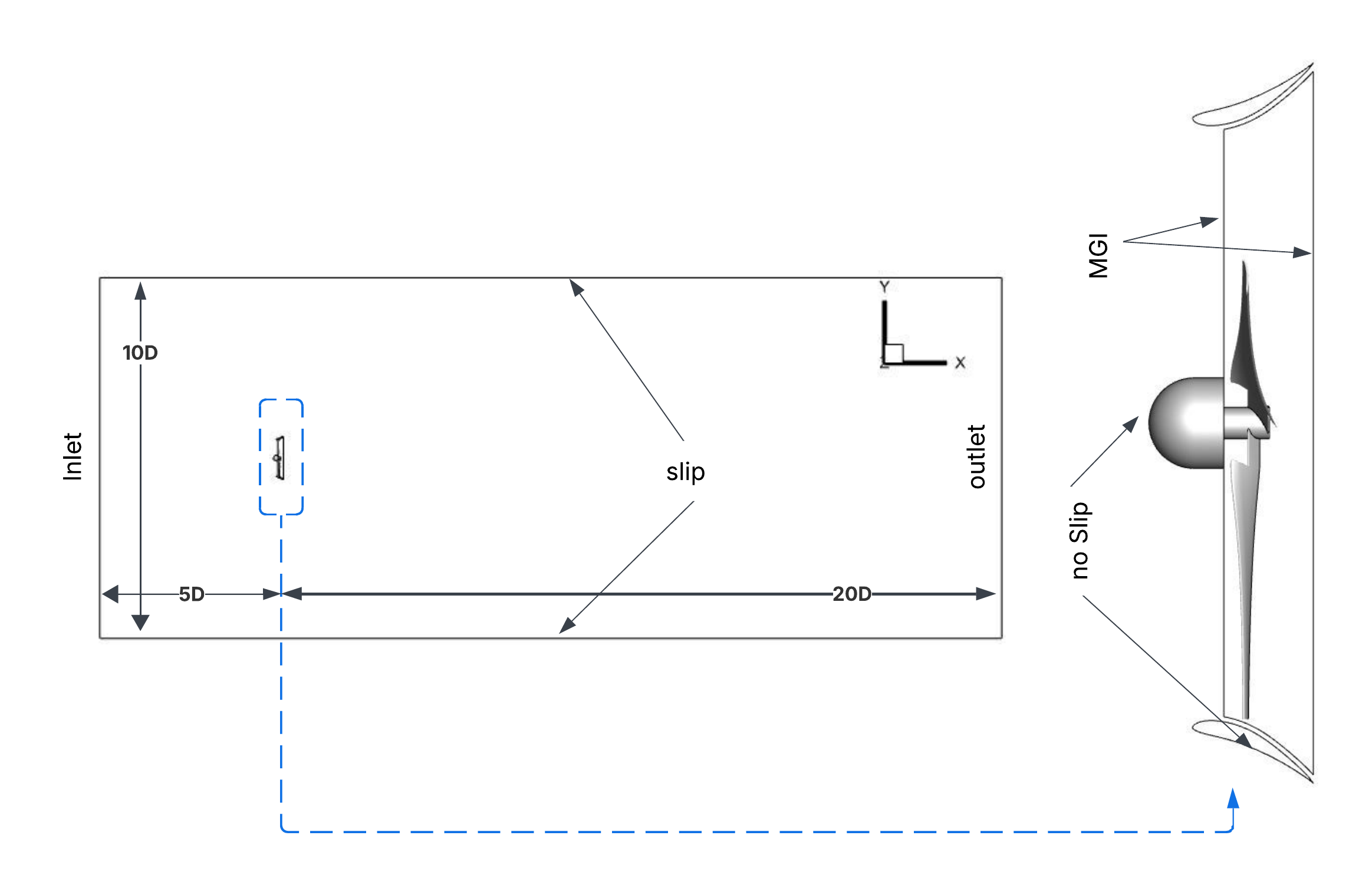}
    \caption{Boundary conditions for the computational domain used for simulating ducted wind turbines}
    \label{bcDWT}
\end{figure}

The boundary conditions implemented in the study are shown in Figure \ref{bcDWT}. Dirichlet inlet and outlet pressure conditions are used for the upstream and downstream boundaries of the fluid domain, respectively, while moving and stationary wall boundary conditions are applied to the rotor and duct, respectively. Slip boundary conditions are used for the outer wall boundaries.

\section{Results}

\subsection{Fluid-Structure Interaction}
The builders of CDWT consistently measured its open rotor to have a Cp of 0.41 \cite{Kanya2018}. The FDR-URANS model employed in this paper predicted a Cp of 0.42 for the open rotor at the optimal tip speed ratio ($\lambda=3.96$). A corresponding thrust coefficient $C_{T}=0.74$ is predicted. 

\begin{figure}[hbt!]
\centering
\includegraphics[width=9cm]{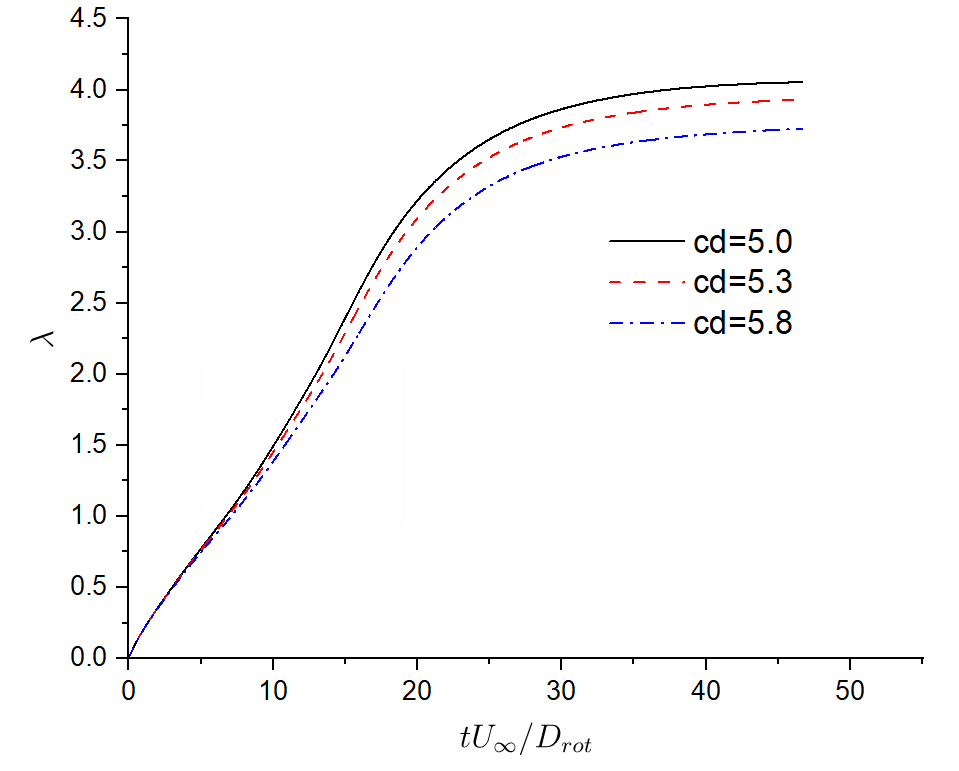}
\caption{The time histories of $\omega_{rot}$ using the FDR-URANS model for the Seyi-Chunlei Ducted Turbine when the damping coefficients are set to 5 kg/s, 5.3 kg/s, and 5.8 kg/s respectively. }
\label{omega_time_hist}
\end{figure}

The FDR model in conjunction with the MGI technique is effective in simulating DWT self-starting and subsequently predicting a fully developed flow characterized by a saturated tip speed ratio. The time histories of $\omega_{rot}$ are plotted in Figure \ref{omega_time_hist} using the FDR-URANS model for the Seyi-Chunlei Ducted Wind Turbine. For different damping coefficients, the SCDT rotor-at-rear configuration reaches equilibrium when the rotor has zero acceleration, the saturated tip speed ratios are 4.05, 3.92, and 3.72 when the damping coefficients $C_d$ are set to 5 kg/s, 5.3 kg/s, and 5.8 kg/s, respectively. 

\subsection{Flow field around the diffuser}

\begin{figure}[ht!]
\centering
\includegraphics[width=14cm]{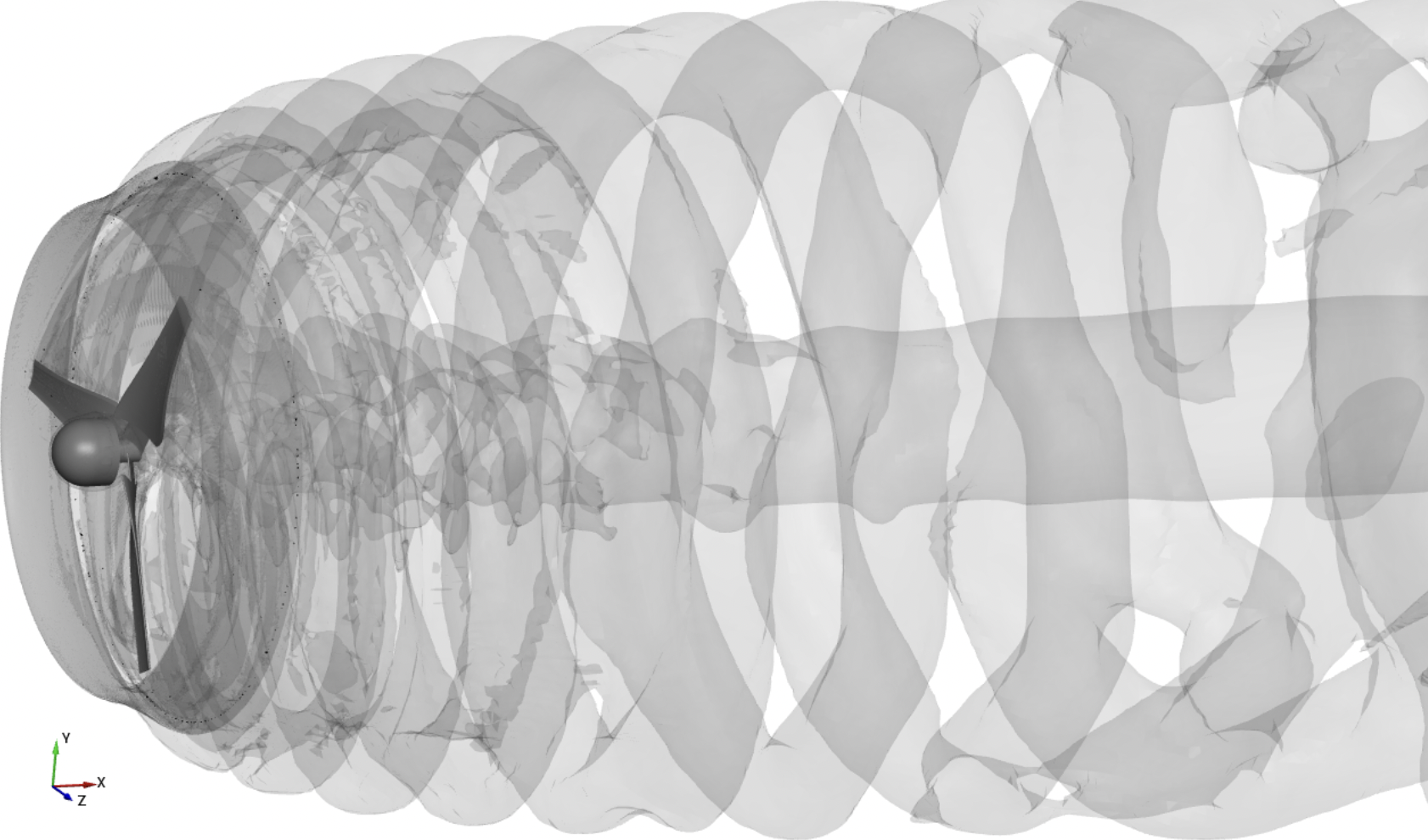}
\caption{The iso-surfaces of Q criteria around the Clarkson Ducted Wind Turbine showing tip vortices and hub vortices. }
\label{Eppler25IsoQ}
\end{figure}

The present results of instantaneous isosurfaces of the Q criteria for the CDWT configuration in Figure \ref{Eppler25IsoQ} show that the FDR model is correctly configured employing a robust mismatched grid interface (MGI) technique. Using this FDR-URANS model of SimericsMP+, our research investigates the aerodynamic performance of both SCDT and CDWT. 

\begin{figure}[ht!]
\centering
\includegraphics[width=12cm]{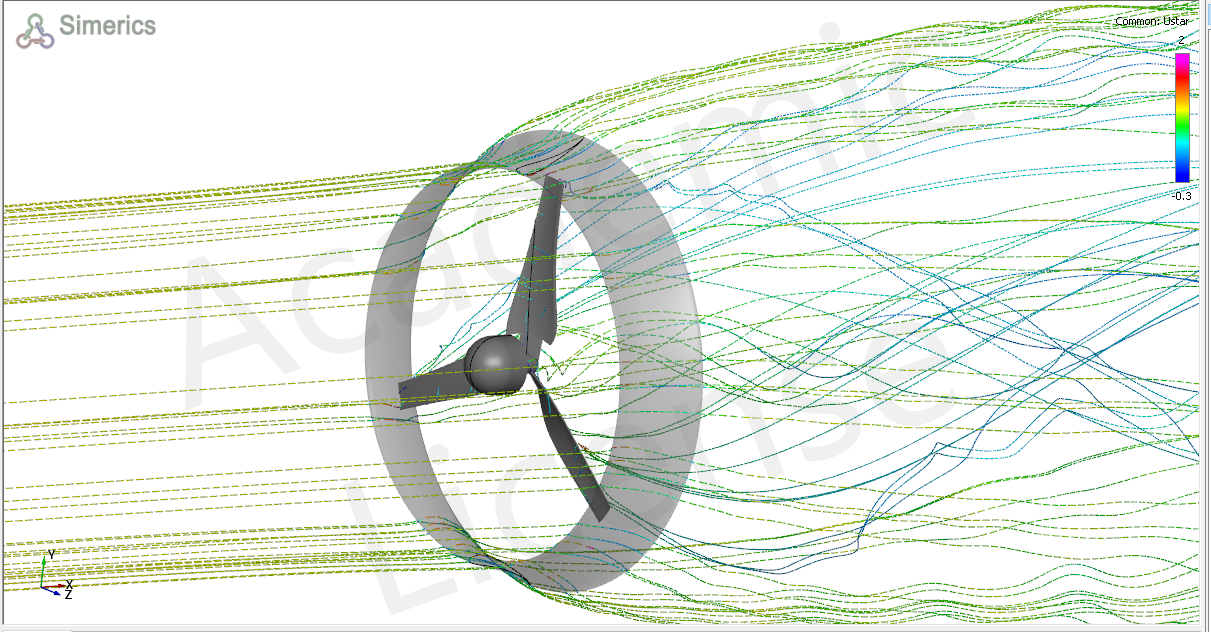}
\caption{Instantaneous velocity streamlines around the Seyi-Chunlei Ducted Turbine at an optimal tip speed ratio.}
\label{SCDT3Dstreamlines}
\end{figure}

Streamlines based on 3D velocity vectors for SCDT are shown in Figure \ref{SCDT3Dstreamlines}. The gap flows between blade tips and the duct are accelerated. The streamlines attached to the duct's inner wall are evident in the figure. The wake flow velocities behind the rotor are significantly below the free-stream velocity due to the extraction of flow kinetic energy to the rotational energy of the blades.

For the Clarkson Ducted Wind Turbine, with the section of Eppler airfoil on the duct at the $25^{o}$ angle of attack, the flow remains largely attached to the inner wall of the duct in the presence of the rotor. However, the CDWT diffuser using the Eppler 423 section airfoil at $25^{o}$ AoA was discovered by Safford et al. \cite{Safford2024} and Ding et al. \cite{Ding2023} to generate flow separation at the trailing edge of the diffuser. 

\begin{figure}[ht!]
\centering
\includegraphics[width=0.46\linewidth]{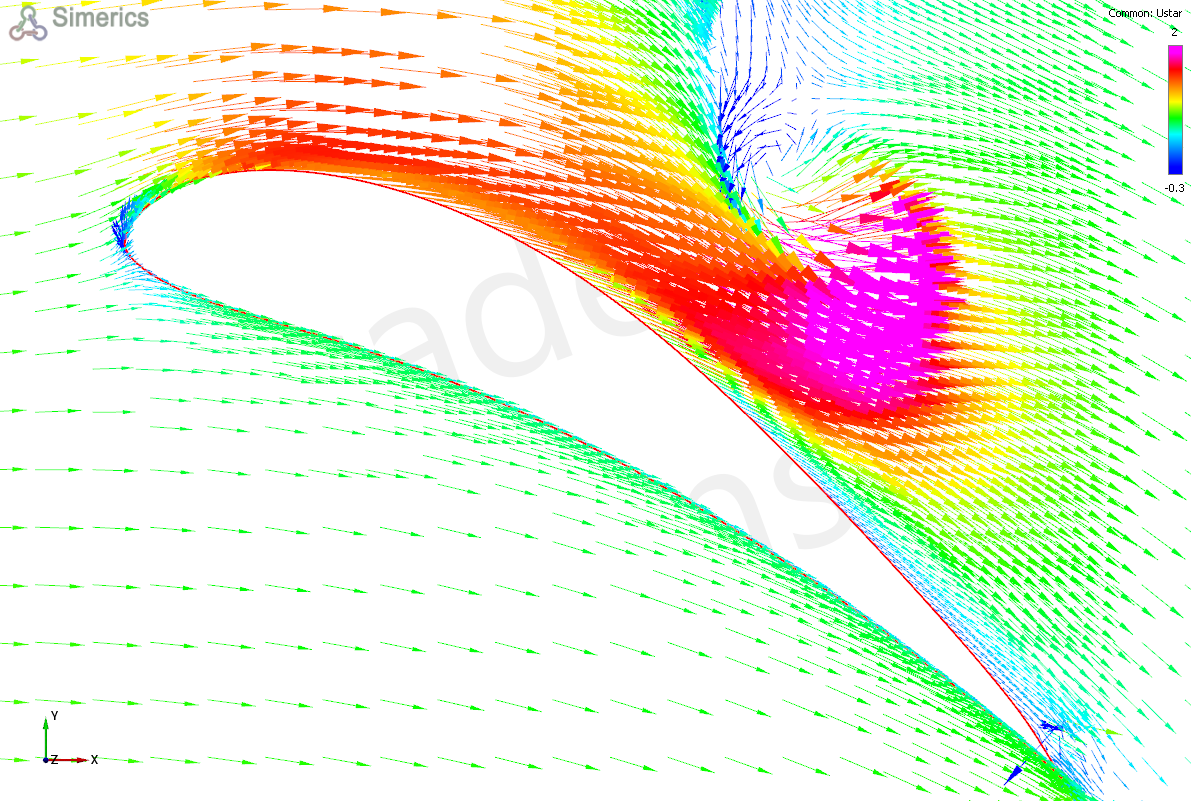} \hfill
\includegraphics[width=0.53\linewidth]{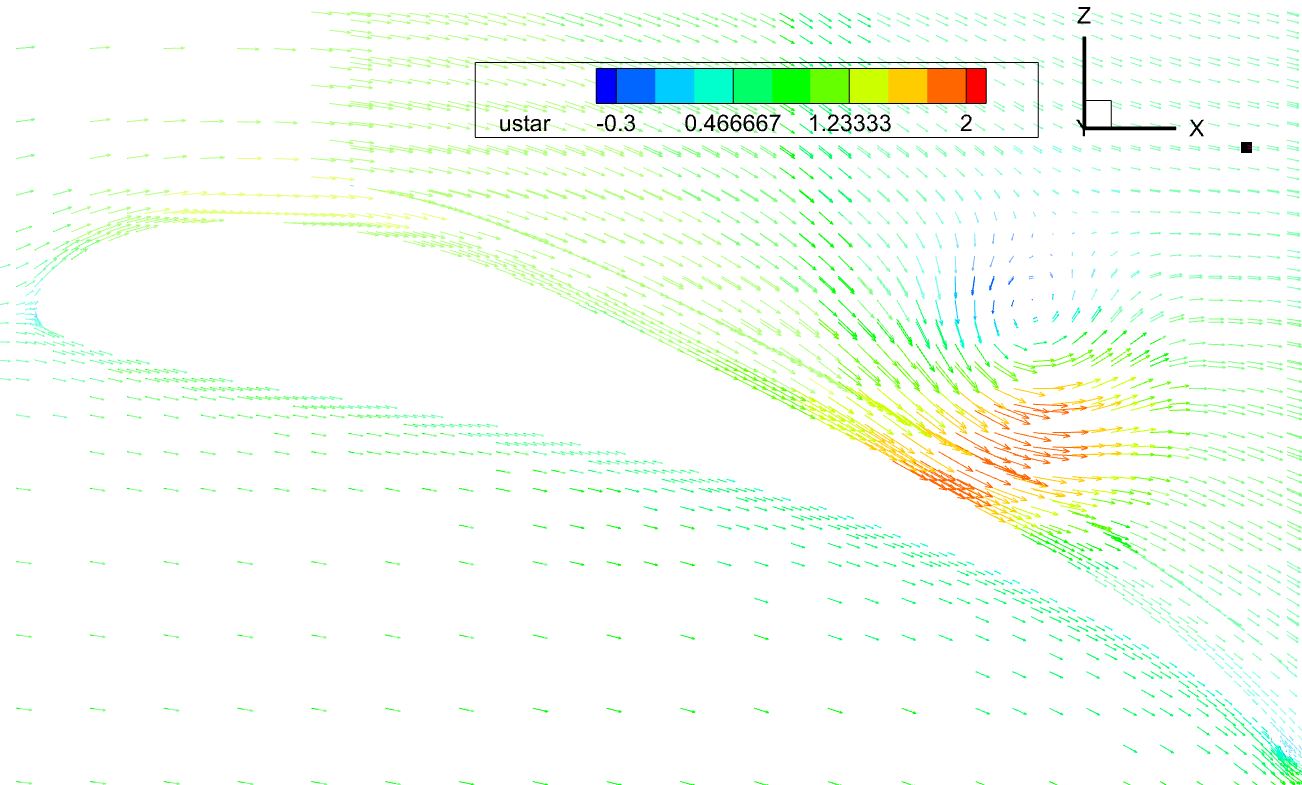} \\
 \hspace{3cm}(a) \hspace{1cm} \hfill (b)  \hspace{2cm}    
\caption{Instantaneous velocity vectors around (a) the Eppler25 diffuser of CDWT and (b) the Selig20 diffuser of SCDT.}
\label{Eppler25vec}
\end{figure}

Our FDR-URANS results suggest that the optimal angle of attack of the Eppler diffuser is overestimated by the lower-fidelity RANS actuator disk model \cite{Bagheri2018,Bagheri2021}. The flow separation can be examined by plotting the gap flow between the blade and the diffuser for both the CDWT and the Seyi-Chunlei Ducted Turbine. The FDR-URANS model is able to predict the interaction between blade tip vortices and the gap flow. The velocity vectors around the Eppler25 diffuser are evidently separated from the inner wall of the duct near the trailing edge of the airfoil, as shown in Figure \ref{Eppler25vec} (a). It is satisfying to confirm that the velocity vectors around the Selig20 diffuser of SCDT remain consistently attached to the inner wall of the duct, as shown in Figure \ref{Eppler25vec} (b). Therefore, one can expect that SCDT is a quieter turbine than CDWT.

\subsection{Power Efficiency of Two Ducted Wind Turbines}

\begin{figure}[H]
    \centering
    \begin{subfigure}[b]{0.49\textwidth}   
        \centering
        \includegraphics[width=\textwidth]{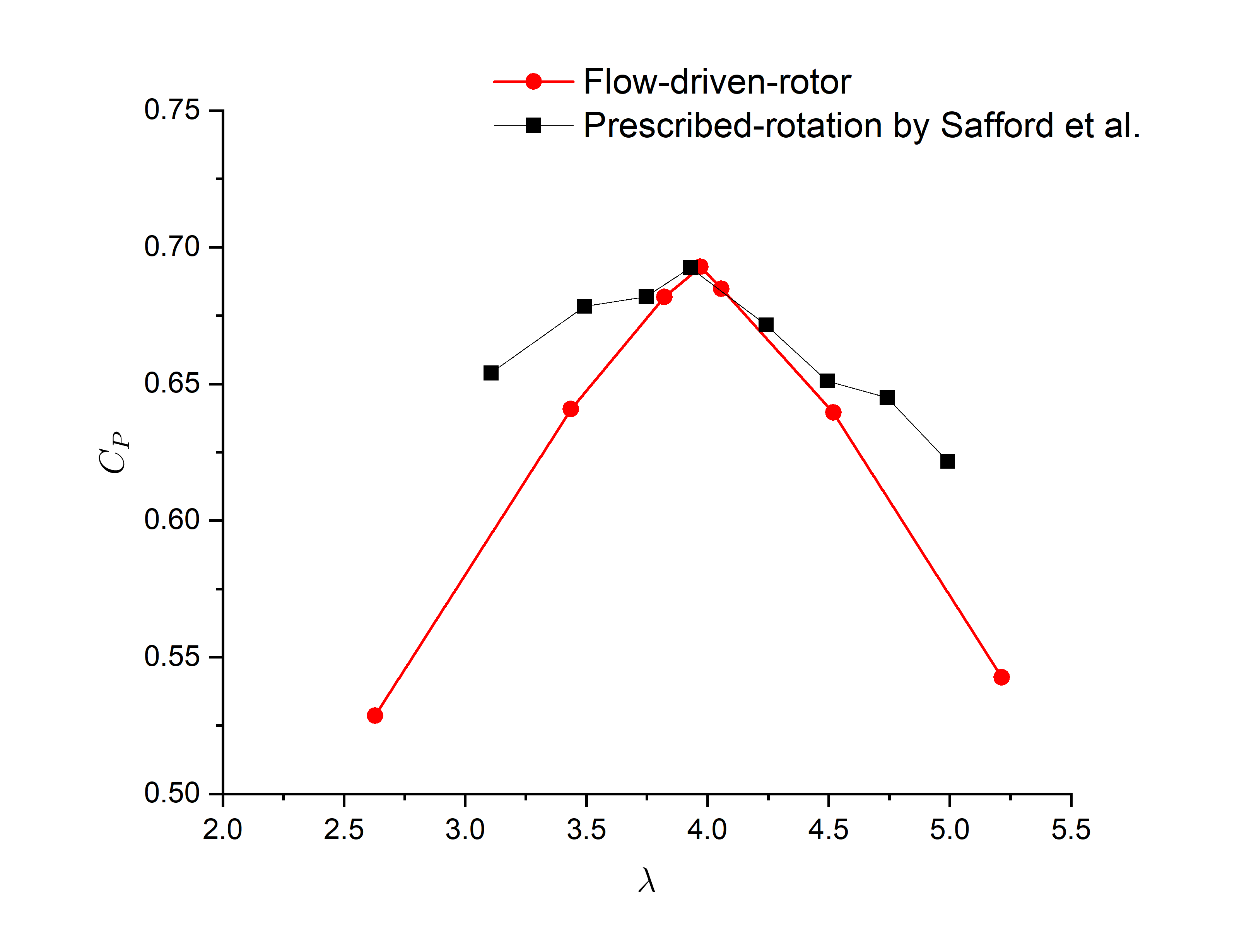}
        \caption{}
    \end{subfigure}
    \hfill
    \begin{subfigure}[b]{0.49\textwidth}   
        \centering
        \includegraphics[width=\textwidth]{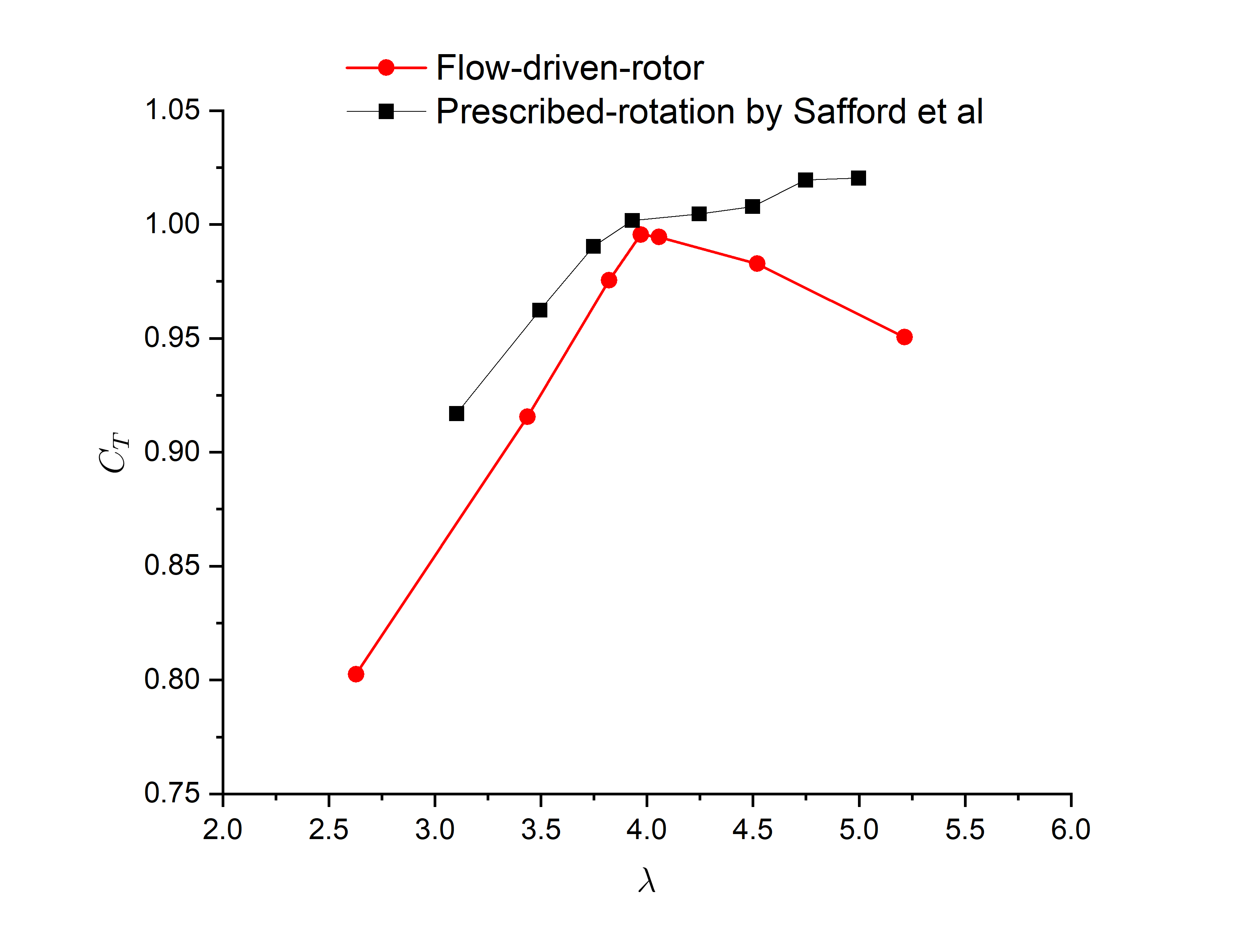}
        \caption{}
    \end{subfigure}
    \caption{$C_p$ and $C_T$ of CDWT at different tip speed ratios.}
    \label{CpEppler25}
\end{figure}

By varying the damping coefficient in the FDR model, we obtain different tip speed ratios. Figure \ref{CpEppler25} (a) plots Cp based on the rotor area obtained for different tip speed ratios of the current study compared to the results obtained from Safford et al. \cite{Safford2024} using a prescribed rotation model. The FDR model suggests the need to be close to the optimal tip speed ratio of 3.93 to maximize the power coefficient. The FDR model is likely more accurate in predicting underperforming ducted wind turbines under ambient conditions when the tip speed ratios are away from $\lambda_{optimal}$. The transient behavior of ambient wind conditions is a key factor in the reduction of Cpt by attenuating the effectiveness of the diffuser in capturing the desired larger mass flow rate \cite{Visser2024}. CDWT has a narrow range of $\lambda$ for peak power production due to the rapid deterioration of Cp even if $\lambda$ is a short distance from the optimal tip speed ratio.

The current FDR model predicts an optimal thrust coefficient of $C_{T}=1$ at the optimal tip speed ratio of 3.93 as shown in Figure \ref{CpEppler25} (b). It should be noted that the prescribed motion model fails to predict the optimal thrust coefficient because $C_{T}$ continues to rise after $\lambda$ surpasses 3.93.  The results of Safford et al. \cite{Safford2024} suggested that the thrust coefficient continues to increase as the tip speed ratio increases further beyond 3.93. Therefore, the FDR model is more reliable in predicting the underperformance for DWT operating at non-optimal tip speed ratios.

\begin{figure}[H]
    \centering
    \begin{subfigure}[b]{0.49\textwidth}   
        \centering
        \includegraphics[width=\textwidth]{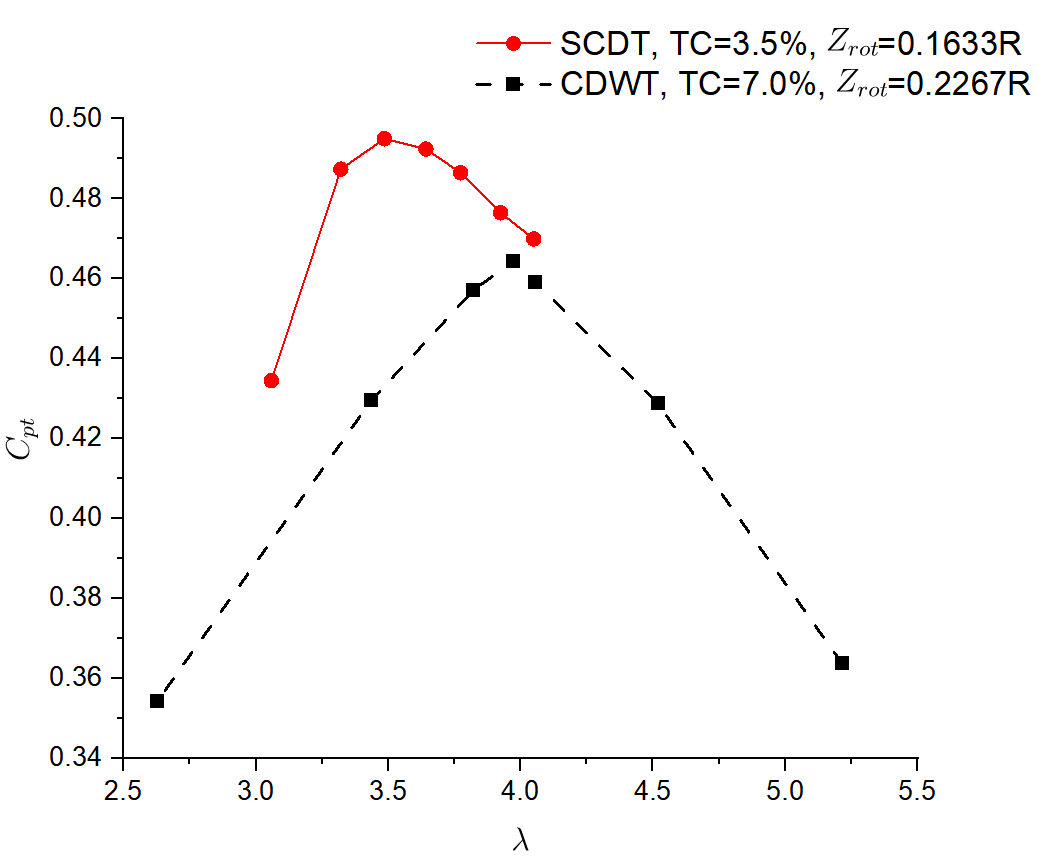}
        \caption{}
    \end{subfigure}
    \hfill
    \begin{subfigure}[b]{0.49\textwidth}   
        \centering
        \includegraphics[width=\textwidth]{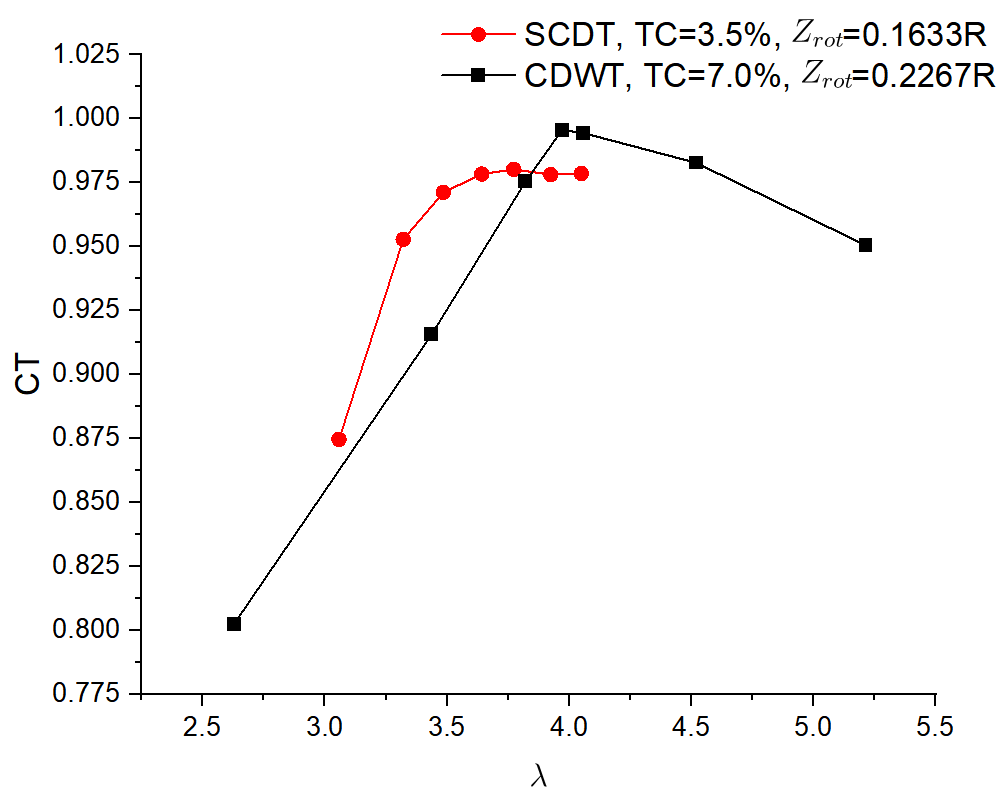}
        \caption{}
    \end{subfigure}
    \caption{(a) Power coefficients  $C_{pt}$ and (b) Thrust coefficients $C_T$ for the Seyi-Chunlei Ducted Turbine at different tip speed ratios}
    \label{CptEppler25SCDT}
\end{figure}

The power coefficient data emphasized in this paper are in terms of Cpt, not Cp, whereby the power is scaled by the rotor area, but instead, the data are scaled by the duct exit area, i.e. the maximum projected area of the duct. The traditional value of Cp would be equal to Cpt multiplied by the ratio of the areas, that is, $Cp = Cpt \times A_{exit}/A_{rotor}$ where $A_{exit}$ is the exit area of the diffuser. Our new design, SCDT, produces a higher Cpt consistently than the Clarkson DWT as shown in Figure \ref{CptEppler25SCDT}. CDWT has an optimal $\lambda$ of 3.97 where the optimal Cpt is 0.464 and the optimal $C_T$ is 0.995. The optimal tip speed ratio for SCDT is 3.49. The corresponding Cpt and $C_T$ for SCDT are 0.495 and 0.97 respectively. This is 6.7\% higher than the optimal Cpt value of 0.464 for CDWT. Moreover, SCDT has a wider range of optimal tip speed ratios, allowing it to harvest more wind energy under ambient wind conditions than the Clarkson DWT. For example, when $\lambda$ is 3.8, the Cpt of SCDT remains as high as 0.48, compared to the CDWT's Cpt of 0.457.  The Eppler25 diffuser is shown to have a pronounced flow separation at its trailing edge. In contrast, the Selig20 diffuser keeps the gap flow attached at its trailing edge. Therefore, SCDT is likely a quieter wind turbine.  

\begin{figure}[H]
    \centering
    \begin{subfigure}[b]{0.49\textwidth}   
        \centering
        \includegraphics[width=\textwidth]{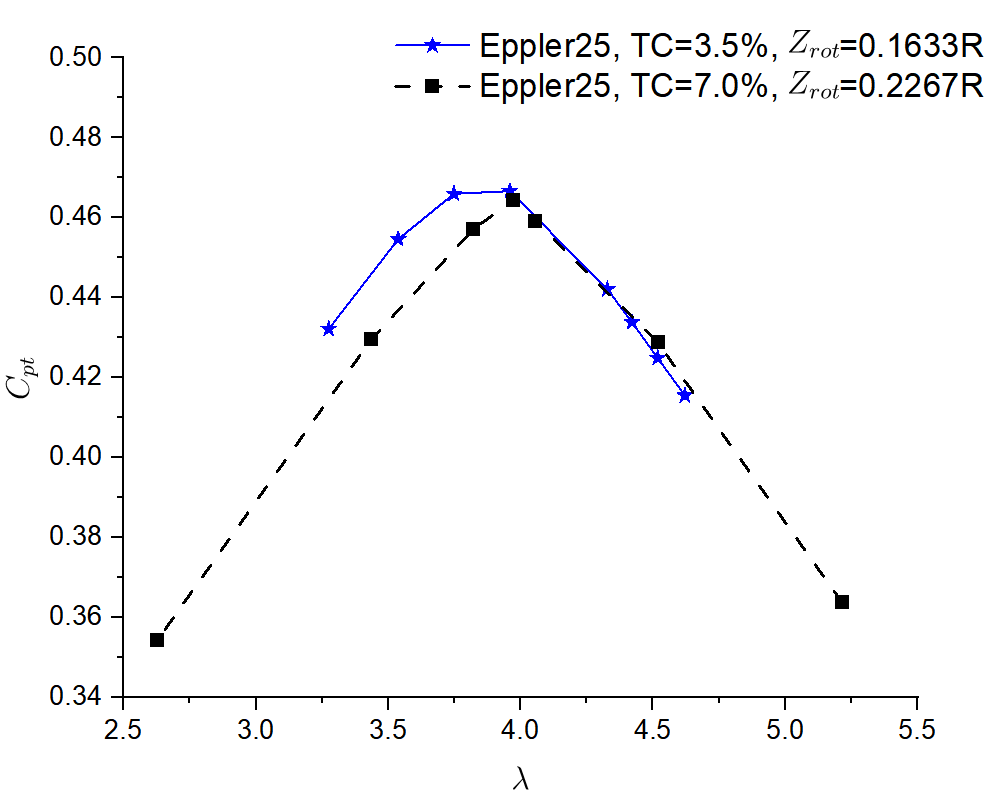}
        \caption{}
    \end{subfigure}
    \hfill
    \begin{subfigure}[b]{0.49\textwidth}   
        \centering
        \includegraphics[width=\textwidth]{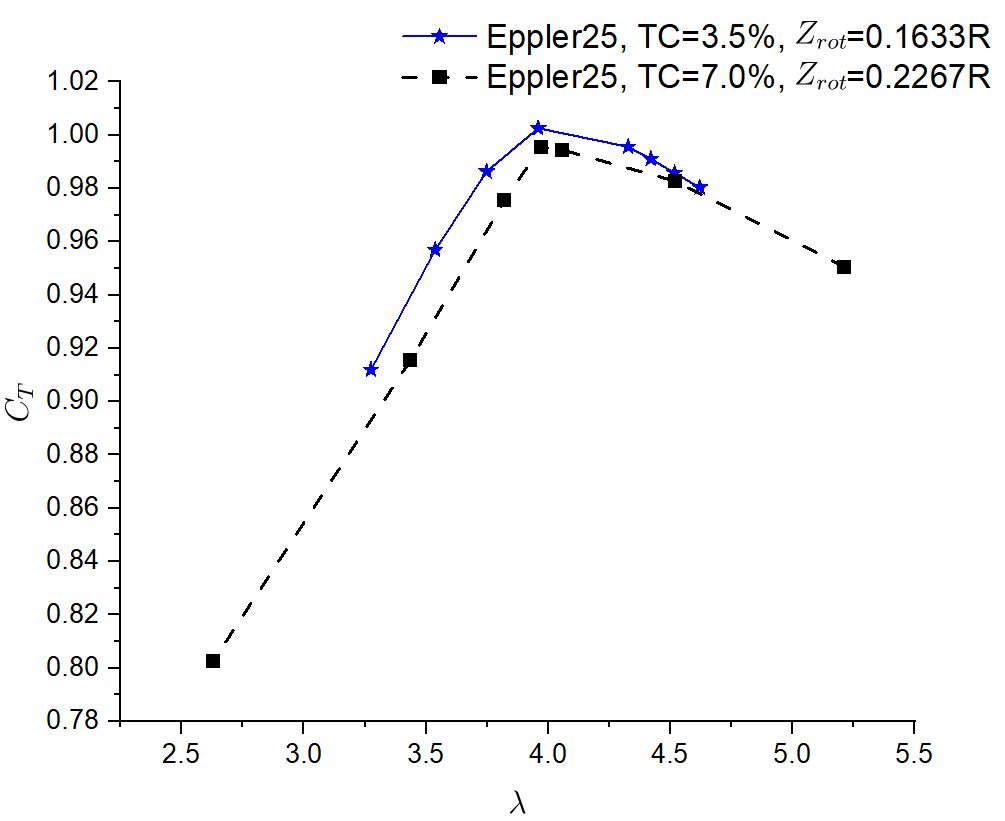}
        \caption{}
    \end{subfigure}
    \caption{(a) Power coefficients $C_{pt}$ and (b) Thrust coefficients $C_T$ for the Clarkson Ducted Wind Turbine with two different tip-speed ratios}
    \label{CptEppler25TC}
\end{figure}

The previously mentioned CDWT design adopted a rather wide tip clearance, i.e. 7\% $R_{rot}$ the same as that adopted by Ding et al. \cite{Ding2023} and Safford et al. \cite{Safford2024}. The FDR model is also used to study the aerodynamic performance of CDWT with narrower tip clearance, that is, 3. 5\%$R_{rot}$. The blade tip clearance ratio is shown to have a marginal impact on both Cpt and $C_T$ as shown in Figure \ref{CptEppler25TC}. The smaller tip clearance leads to slightly higher thrust coefficients overall as well as slightly higher Cpt at tip speed ratios below the optimal value.   

\newpage

\begin{figure}[H]
    \centering
    \begin{subfigure}[b]{0.49\textwidth}   
        \centering
        \includegraphics[width=\textwidth]{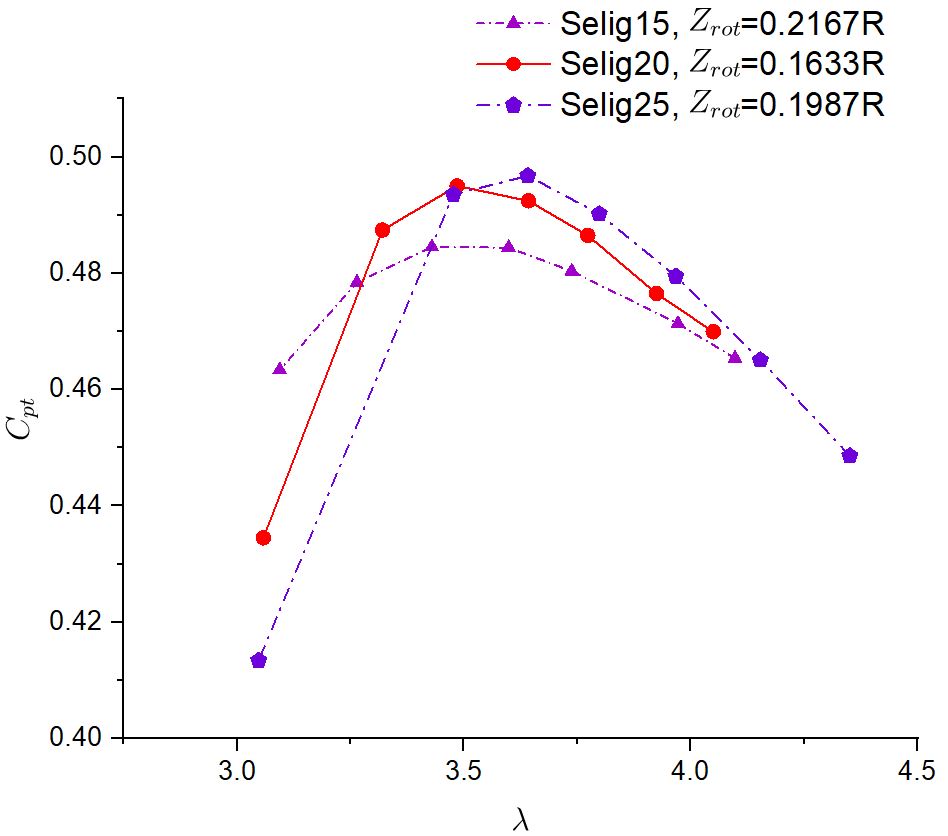}
        \caption{}
    \end{subfigure}
    \hfill
    \begin{subfigure}[b]{0.49\textwidth}   
        \centering
        \includegraphics[width=\textwidth]{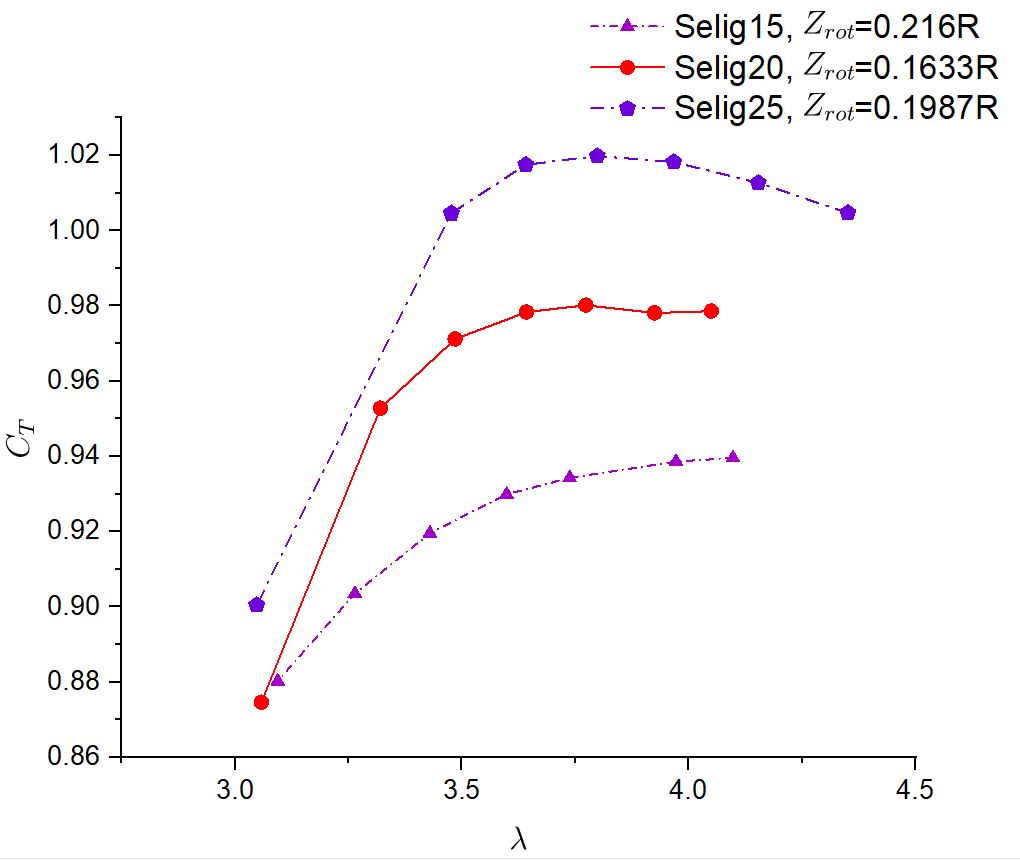}
        \caption{}
    \end{subfigure}
    \caption{(a) Power coefficients $C_{pt}$ and (b) Thrust coefficients $C_T$ for diffusers with Selig 1223 airfoil}
    \label{allselig}
\end{figure}

Table \ref{tab:optimal table} lists the optimal tip speed ratio and $C_{pt}$ in the various design configurations. It is shown that among the various configurations, the Seyi-Chunlei Ducgted Turbine (Selig20) is very competitive and gives the best performance. To achieve the same $C_{pt}$ in Selig25, the rotor must spin faster, which suggests that SCDT is the best choice among all the designs listed for ambient wind conditions. In addition, Figure \ref{allselig} (a) shows that the performance of the Seyi-Chunlei duct turbine is close to Selig25 while we expect more acoustic signature and flow separation in the Selig25 design. Figure \ref{allselig} (b) confirms that the Selig25 design requires larger thrust coefficients to achieve optimal Cpt.

\begin{table}[hbt!]
\centering
\caption{The optimal  values of both Eppler 423 and Selig 1223 }
\label{tab:optimal table}
    \begin{tabular}{ p{2cm}|p{2cm}|p{2cm}|p{1.5cm}|p{1cm}|p{1cm}  }
        \hline
        \multicolumn{6}{c}{Optimal Parameters} \\
        \hline
        Airfoil Name& Attack Angle (degree) & Tip Clearance (in $\%$ of $R_b$) & Tip Speed ($\lambda$) & $C_{pt}$ & $C_{T}$  \\
        \hline
        Eppler 423 & 25 & 7.0 &3.97 & 0.464 &1.00 \\
        Eppler 423 & 25 & 3.5 &3.96 & 0.467 &1.00 \\
        Selig 1223 & 15 & 3.5 &3.43 & 0.485 &0.92 \\
        Selig 1223 & 20 & 3.5 &3.49 & 0.495 &0.971 \\
        Selig 1223 & 25 & 3.5 &3.64 & 0.497 &1.02 \\      
        \hline
    \end{tabular}
\end{table}

\section{Conclusions}

In this paper, the flow-driven-rotor (FDR) and URANS models of Simerics-MP + are employed to successfully simulate a ducted wind turbine self-starting and predict a fully developed flow for a saturated tip speed ratio. The FDR model successfully predicts the optimal thrust coefficient, whereas the prescribed rotation model fails to do so. Although the optimal Cp predicted by the FDR model is fairly close to the prediction of the prescribed motion model \cite{Safford2024}, the FDR model is generally more accurate in predicting underperformance under ambient wind conditions when the tip speed ratio is non-optimal. The FDR model does offer a new route for simulating ducted wind turbines in ambient wind conditions. In addition, in this paper an improved ducted wind turbine design, the Seyi-Chunlei Ducted Turbine (SCDT), is proposed. SCDT keeps gap flows strongly attached to the inner wall of the diffuser and has considerably less flow separation at the duct's trailing edge than the Clarkson Generation I ducted wind turbine. The effectiveness of SCDT is further verified because the tip vortices are pushed to the inner wall of the duct and propagate downstream along the gap flow streamlines. SCDT gives approximately 7\% higher peak Cpt than the Clarkson Ducted Wind Turbine. Moreover, SCDT has a wider range of optimal tip speed ratios, helping it to harvest more wind energy under ambient conditions. SCDT is likely a much quieter turbine in comparison to the Clarkson Ducted Wind Turbine. Because the computational analyses in this paper are based on incompressible flow solutions, SCDT is also expected to be a competitive ducted water turbine operating at similar Reynolds numbers. 

\section{Acknowledgement}
The second author thanks a Clarkson Ignite Graduate Fellowship awarded to him between Spring 2020 and Spring 2025. Three MS students have been supported by this grant from Clarkson University. The authors thank Junfeng Wang and Joel Varghese of Simerics Inc. for the software license renewal and for their assistance in some of our computations. 

 \bibliographystyle{new-aiaa} 
 \bibliography{DWT_references}

\end{document}